\documentclass[conference]{IEEEtran}
\IEEEoverridecommandlockouts
\usepackage{cite}
\usepackage{amsmath,amssymb,amsfonts}
\usepackage{algorithmic}
\usepackage{graphicx}
\usepackage{textcomp}
\usepackage{xcolor}
\usepackage{listings}
\usepackage{booktabs}
\usepackage{fontawesome}
\usepackage{pifont}
\usepackage{amsmath}
\usepackage{enumitem}
\usepackage{soul}
\usepackage{tabularx, multirow}
\usepackage[hidelinks]{hyperref}

\usepackage{subfigure}
\usepackage{tikz}
\definecolor{codegreen}{rgb}{0,0.6,0}
\definecolor{codegray}{rgb}{0.5,0.5,0.5}
\definecolor{codepurple}{rgb}{0.58,0,0.82}
\definecolor{backcolour}{rgb}{0.95,0.95,0.92}

\lstdefinestyle{customc}{
	commentstyle=\color{codegreen},
	keywordstyle=\color{magenta},
	numberstyle=\tiny\color{codegray},
	stringstyle=\color{codepurple},
	basicstyle=\ttfamily\scriptsize,
	breakatwhitespace=false,         
	breaklines=true,                 
	captionpos=b,                    
	keepspaces=true,                 
	numbers=left,                    
	numbersep=3pt,                  
	showspaces=false,                
	showstringspaces=false,
	showtabs=false,                  
	tabsize=2,
	language=C,
	xleftmargin=0.3cm
}

\definecolor{redish}{RGB}{255,204,204}
\definecolor{greenish}{RGB}{61,102,95}
\newcommand*\circlered[1]{\tikz[baseline=(char.base)]{%
		\node[white,shape=circle,fill=redish,draw,inner sep=1pt] (char) {\color{black}\sffamily #1};}}

\newcommand*\circlegreen[1]{\tikz[baseline=(char.base)]{%
		\node[white,shape=circle,fill=greenish,draw,inner sep=1pt] (char) {\color{white}\sffamily #1};}}

\author{
    \IEEEauthorblockN{Salvatore Di Girolamo, Daniele De Sensi, Konstantin Taranov, Milos Malesevic,\\Maciej Besta, Timo Schneider, Severin Kistler, Torsten Hoefler}
    \IEEEauthorblockA{Dept. of Computer Science, ETH Z{\"u}rich, Switzerland
    \\\{first-name.last-name\}@inf.ethz.ch}
}

\definecolor{r1}{RGB}{87,114,158}
\definecolor{r2}{RGB}{204,137,99}
\definecolor{r3}{RGB}{93,157,107}
\definecolor{r4}{RGB}{196,78,82}
\definecolor{r5}{RGB}{129,114,180}
\usepackage{marginnote}
\usepackage{soul} %
\usepackage{hhline}
\definecolor{lightyellow}{RGB}{250, 250, 180}
\definecolor{HLYELLOW}{RGB}{240, 127, 0}
\definecolor{hlyellow}{RGB}{240, 127, 0}
\sethlcolor{lightyellow}
\definecolor{lightcyan}{RGB}{160,255,255}
\usepackage{xparse}

\usepackage{mdframed}
\global\mdfdefinestyle{review}{%
	linecolor=lightyellow,linewidth=3pt,%
	leftmargin=0cm,rightmargin=0cm,%
	skipabove=0cm,skipbelow=0cm,%
	innerrightmargin=0cm,innerleftmargin=0cm,%
	innerbottommargin=0cm,innertopmargin=0cm,%
	backgroundcolor=lightyellow
}
\global\mdfdefinestyle{reviewtext}{%
	linecolor=lightyellow,linewidth=0pt,%
	leftmargin=0cm,rightmargin=0cm,%
	skipabove=0.1cm,skipbelow=0.1cm,%
	innerrightmargin=0cm,innerleftmargin=0cm,%
	innerbottommargin=0cm,innertopmargin=0cm,%
	backgroundcolor=lightyellow
}

\begin{document}

\newif\ifrev

\ifrev
\DeclareDocumentCommand\review{m g g}{%
	{\IfNoValueF {#2}{%
			\IfNoValueF {#3}{%
				{\marginnote{\sethlcolor{#3}\hl{\normalfont \textbf{{\normalsize{\color{white}#2}}}}}%
				}%
			}%
			\IfNoValueT {#3}{%
				{\marginnote{\normalfont \textbf{\normalsize{#2}}}%
				}%
			}%
		}%
		\hl{#1}%
	}%
}
\else
\DeclareDocumentCommand\review{m g g}{#1}
\fi

\title{\vspace{-0.5em}Building Blocks for Network-Accelerated \\Distributed File Systems\vspace{-0em}}

\maketitle
\begin{abstract}
High-performance clusters and datacenters pose increasingly demanding requirements on storage systems. If these systems do not operate at scale, applications are doomed to become I/O bound and waste compute cycles. 
To accelerate the data path to remote storage nodes, remote direct memory access (RDMA) has been embraced by storage systems to let data flow from the network to storage targets, reducing overall latency and CPU utilization. 
Yet, this approach still involves CPUs on the data path to enforce storage policies such as authentication, replication, and erasure coding. 
We show how storage policies can be offloaded to fully programmable SmartNICs, without involving host CPUs. 
By using PsPIN, an open-hardware SmartNIC, we show latency improvements for writes (up to 2x), data replication (up to 2x), and erasure coding (up to 2x), when compared to respective CPU- and RDMA-based alternatives.
\end{abstract}

\section{Introduction}
Distributed File Systems (DFSs) play a fundamental role in tackling the growing I/O bottleneck. By decoupling control and data planes, these architectures can be easily managed and scaled out. 
While there exist a plethora of DFS architectures, it is possible to identify building blocks that are fundamental and typically implemented by all of them. 
For example, clients must be authenticated and their requests must be validated. If this does not happen, a client can write to any storage location, violating tenant isolation and potentially bringing the file system to an inconsistent state. 
Additionally, data must be stored resiliently by either replicating it on different storage nodes or storing it together with parity blocks (i.e., erasure-coded). Without resiliency, the failure of a single storage node can compromise the entire file system or large parts of it.

These building blocks, which we call \emph{DFS policies}, are defined by the DFS control plane and enforced in the data plane. For example, file or object metadata store access permissions and whether and how the file or object must be replicated or erasure-coded. Whenever clients access the data, these policies must be enforced: i.e., the client request must be validated and the data must be eventually replicated or erasure-coded. 

Until recently, the performance of data plane storage operations has been greatly limited by storage media performance. Hence, factors like network overheads, data copies, and CPU utilization were of negligible importance.
This led to the introduction of complex software layers into the storage nodes to implement strategies for easing storage media bottlenecks (e.g., batching, striping), and enforce DFS policies.
 
However, while this assumption does not hold at all for in-memory file systems, it must also be revised for persistent storage. In fact, with the emergence of dense, byte-addressable non-volatile main memories (NVMMs)~\cite{sainio2016nvdimm} and NVMe JBOFs (Just a Bunch of Flash)~\cite{min2021gimbal}, storage media are able to ingest data at network speed or faster. Hence, network performance and software overheads start playing an important role and must be optimized to not let them become bottlenecks.
In this direction, remote direct memory access (RDMA)~\cite{infiniband2004infiniband} has been the focus of many DFS optimizations~\cite{braam2019lustre, schmuck2002gpfs, borthakur2008hdfs, wang2014parallel, ross2000pvfs}. 
RDMA provides low latency and high bandwidth one-sided communications that allow hosts to access memory of remote peers (e.g., remote storage nodes) without involving their CPUs.

Unfortunately, RDMA provides negligible compute capabilities, which are insufficient to express custom DFS policies.
Consequently, DFSs usually still rely on storage node CPUs to enforce custom policies by using remote procedure calls (RPC) for triggering policy enforcement and RDMA to move data. However, this approach loses the one-sided characteristic of RDMA, which is key for getting low latency and high throughput. To work around this issue and fully embrace RDMA, some DFSs~\cite{yang2019orion, lu2017octopus, yang2020filemr} delegate policy enforcement to clients. However, this can lead to worse performance (e.g., for replication, a client issues multiple writes, one for each replica) or require higher trust (e.g., accesses from fully trusted clients do not require validation).

\begin{figure*}[t!]
	\centering
	\vspace{-1em}
	\subfigure[]{\includegraphics[width=0.27\textwidth]{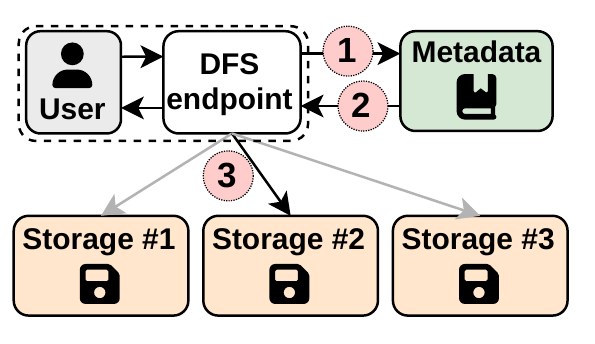}} 
	\subfigure[]{\includegraphics[width=0.23\textwidth]{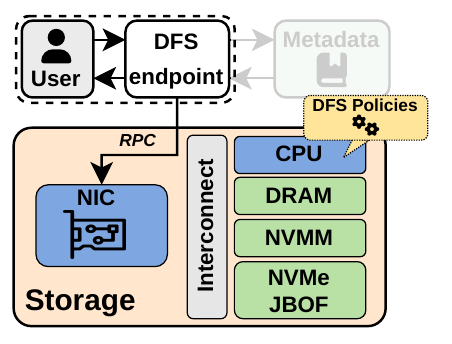}} 
	\subfigure[]{\includegraphics[width=0.23\textwidth]{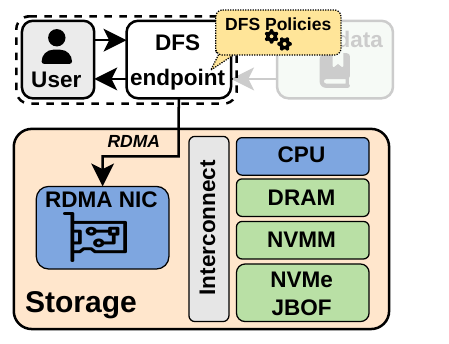}}
	\subfigure[]{\includegraphics[width=0.23\textwidth]{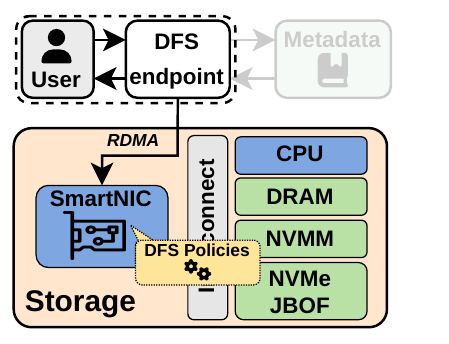}}
	\vspace{-0.9em}
	\caption{(a) DFS operational model assumed in this paper; (b) DFS policies implemented in storage nodes CPU; (c) with RDMA; and (d) with SmartNICs.}
	\label{fig:overview}
	\vspace{-1.5em}
\end{figure*}

We show how DFS policy enforcement can be offloaded to SmartNICs (smart network interface cards). When offloaded, these policies are enforced directly on the SmartNIC, without involving the storage node CPU.
Additionally, we show how packet-level processing techniques can accelerate the execution of these policies, even when compared to the case where they are enforced by the CPU of the storage nodes. 
\emph{This work aims at filling the gap between full-RDMA DFSs, which provide best performance for raw writes but do not support custom DFS policies, and CPU-based DFSs, that allow expressing any DFS policy at the cost of additional overheads (e.g., PCIe latency).}
All in all, we introduce the following contributions:
\begin{itemize}[noitemsep,topsep=1pt,parsep=0pt,partopsep=0pt,leftmargin=10pt]
	\item Show how SmartNICs enable high-performance implementation of DFS-specific protocols;
	\item Demonstrate how fully programmable SmartNICs with packet-level processing capabilities can accelerate multi-node communications required for, e.g., data replication;
	\item Apply packet-processing techniques to offload complex tasks such as erasure coding. Not only this approach results better than proprietary, firmware-based solutions but also improves the time-to-market of new strategies as they are not constrained anymore by vendor-dependent deployments.
	\item We study the benefits of network acceleration for DFSs on an open-hardware, fully programmable SmartNIC. The entire toolchain, comprised of cycle-accurate simulations, is available online, easing reproducibility and enabling quick investigation of new hardware features for future research.
\end{itemize}
We envision that the approach taken in this work to offload DFS building blocks can be followed for designing next-generation DFSs, allowing them to benefit from RDMA performance while effectively enforcing storage policies.

\section{Distributed File Systems and SmartNICs}
\label{sec:dfs_intro}

By decoupling control and data planes, DFSs can distribute I/O accesses and abstract technology- and vendor-specific storage backends, providing an independent and standard interface to the clients while easing the management of conventional storage systems. 
Commonly, DFSs are composed of three main services: management, metadata, and storage. 
Activities such as authentication and monitoring are carried out by the management service.
Control plane tasks are performed by the metadata service that indexes files/objects and references the actual data stored by the storage service. 

Figure~\ref{fig:overview}(a) shows a typical DFS workflow. A user is interfaced with a DFS endpoint that can be either a library, a kernel module, or a separate network node. In the following, we do not make a distinction between users and DFS endpoints, referring to both of them as \mbox{\emph{client}}. The client first authenticates through the DFS endpoint with the management service (not shown in the figure). Then, to access file or object data, it queries the metadata service~\circlered{1} to retrieve the file layout~\circlered{2}. A file layout describes the regions (e.g., objects or blocks) composing a file and address information (e.g., on which storage node a region is stored and at which storage address).
This information allows the client to communicate directly with the storage node for accessing the data~\circlered{3}. 

\subsection{DFS policies} A DFS policy is a set of actions defined by a distributed file system to be enforced when clients access data. These policies are defined in the control plane (i.e., management and metadata service) and enforced in the data plane (i.e., storage nodes). Normally, overheads introduced by the enforcement of these policies are factored in the data access (e.g., read or write) latencies. We investigate how DFS policies can be enforced directly by the network interfaces of the storage nodes. By enforcing them on the NIC, on per-packet basis, we can improve the overall latency of write operations besides lowering CPU usage on the storage nodes.
While each DFS can define its own custom policies, we identify three classes of policies of general interest for DFSs: protocol, data movement, and data processing policies. In this paper, we select one representative policy for each class:

\noindent\textbf{Protocol: client request authentication.}
Data access requests issues by clients to storage nodes must be validated. This validation can avoid the case of a malfunctioning or malicious application acting as a client and bringing the DFS to an inconsistent state. 
With this policy, a client must first obtain a ticket or capability from the metadata node and then use it to make requests to the storage nodes. Storage nodes can verify the legitimacy of the request through the capability.

\noindent\textbf{Data movement: replication.}
Data replication improves reliability and fault tolerance capability of DFSs. The idea is to replicate the data on $k$ storage nodes, where $k$ can be a global, per-pool, or per-file parameter. The metadata nodes store the replication information, and the replication strategy is triggered whenever new data is stored. 

\noindent\textbf{Data processing: erasure coding.}
Data replication is characterized by high storage costs, which are linear in the replication factor.
With erasure coding (EC), data is split into $k$ chunks and stored together with $m$ parity chunks. 
In case of failure, missing chunks can be recovered by using the remaining ones.

Figures~\ref{fig:overview}(b-d) sketch different DFS architectures for the storage nodes. 
In a CPU-centric storage node architecture (Figure~\ref{fig:overview}b), the client issues remote procedure calls (RPC) to the storage node, which trigger software components on the CPU of the storage node enforcing DFS policies. 
Figure~\ref{fig:overview}c shows an RDMA-centric approach, where the CPU of the storage node is bypassed. With this approach, the clients must enforce DFS policies, as RDMA does not expose a programmable interface.
Finally, Figure~\ref{fig:overview}d, shows a SmartNIC-centric approach, which we explore in this paper, where DFS policies are implemented directly in the network interface of the storage nodes.

\subsection{Network acceleration}\label{sec:principles}
As our goal is to offload the enforcement of these policies to the network, we now identify a set of principles that an in-network compute solution should provide in order to enable effective DFS policy offloading.

\noindent\textbf{One-sided requests.}
Storage nodes should handle data accesses and relative policy enforcement without involving the host CPU of the storage nodes and without requiring long-lived per-client data structures.
For example, if storage nodes actively participate in the data replication process (e.g., as they are arranged in a tree or ring virtual topology), the forwarding to next replicas should not involve the host CPU nor require to maintain long-term information about the virtual topology.
While RDMA-based solutions allow one-sided data access, they are not designed to execute custom compute tasks (i.e., DFS policies). 

\noindent\textbf{Flexibility.}
DFS policies can be complex and data-dependent. For example, erasure coding policies need to access the entire packet payload to compute parities. 
For this, RDMA-based solutions would need to rely on vendor-specific accelerators for erasure coding~\cite{shi2019triec}. 
Solutions based on P4~\cite{bosshart2014p4} do not provide iteration constructs (making deep-packet inspection challenging), pointers, references, and inter-packet state.
Solutions based on eBPF/XDP~\cite{ebpf-netronome} are limited by the programming model (e.g., bounded loops, limited number of instructions, limited packet forwarding capabilities)~\cite{miano2018creating}.
While it is possible to extend existing NIC designs~\mbox{\cite{lockwood2007netfpga, forencich2020corundum}} to implement DFS policies in hardware, this would increase the complexity of expressing and deploying such policies.

\noindent\textbf{User-level.} 
DFSs should be able to install custom policies without needing privileged access to the system (i.e., admin rights). This capability opens up network acceleration also to user-level I/O libraries such as DAOS~\cite{hennecke2020daos}.
Expressing policies in eBPF/XDP would mean that the DFS-policy sees all packets coming through the network interface and reacts to specific ones (e.g., write requests). However, this rules out the possibility of having user-space applications (without elevated privileges) installing new policies because it would break isolation principles (i.e., a policy installed by an application could access packets targeting other applications). Similar issues arise for DPDK~\cite{dpdk} and SmartNICs providing programming models based on eBPF/XDP~\cite{ebpf-netronome} and DPDK~\cite{liu2021performance}.

\subsubsection{Streaming processing in the network}
Network communications, or messages, consist of streams of packets traveling between pairs of endpoints. With message-based processing, the receiving endpoint receives the full stream of packets before processing it. This is what happens when processing data received with, e.g., MPI point-to-point communications. With this approach, the NIC has to copy packet payloads to host memory before triggering CPU processing.
A different approach is to leverage the streaming nature of network communications and process packets as the endpoints receive them. This approach exposes packet-level parallelism, potentially accelerating the processing of incoming data as multiple packets are processed in parallel. When data processing is offloaded to the NIC, streaming processing also allows to have lighter memory requirements: i.e., there is no need to buffer the full message on the NIC before being able to process it, but only the packets currently being processed.

sPIN~\mbox{\cite{hoefler2017spin}} is an in-network compute solution operating on packet streams , which enables the offloading of DFS policies according to the above-described principles.
In particular, sPIN enables applications to define lightweight kernels executed on incoming network packets directly on the NIC. These kernels, called \emph{handlers}, are defined on classes of messages. When packets of a message matching a given class arrive, sPIN schedules the corresponding handlers to be executed on the on-NIC \textit{Handler Processing Units} (HPUs). To process packets, an application defines three handlers: the header handler (HH), executed only on the first packet of a message; the completion handler (CH), run on the last packet of a message (i.e., completion packet); and the payload handler (PH), executed on all packets of a message (included header and completion). sPIN requires that the network delivers the header packet first and the completion packet last, without introducing additional constraints on payload packets.

sPIN provides an RDMA+X programming model, where the X is a per-packet task defined by the application running on the host and executed on the NIC,
fully implementing the \mbox{\textbf{one-sided requests}} principle. 
sPIN handlers can be expressed in C/C++ and, differently from P4 and eBPF/XDP, are not subject to restrictions on actions that can be performed on the incoming packets, providing \mbox{\textbf{flexibility}}. 
Finally, differently from solutions based on DPDK, P4, and eBPF/XDP, sPIN enables \mbox{\textbf{user-level}} applications to offload compute tasks to the NIC, providing isolation by design. This is achieved by matching packets to application-defined execution contexts (i.e., providing information on, e.g., which packet handlers to run), similarly to the way packets are matched to queue pairs in RDMA or to match list entries in Portals 4~\mbox{\cite{portals42}}.

In this work, we use an open-source implementation of sPIN, PsPIN~\cite{pspin}. PsPIN is a PULP-based~\cite{pulp} packet processor composed of 32 RISC-V cores running at 1 GHz (i.e., the HPUs), divided into four compute clusters. Each compute cluster is equipped with a 1 MiB single-cycle access memory (L1) and an off-cluster 4 MiB memory (L2). In addition, the accelerator includes a hardware packet scheduler that provides low-latency scheduling (1-2 cycles) and DMA engines to move data between NIC memories and interface with host memory.

\begin{figure}[t]
	\centering
	\includegraphics[width=\columnwidth, trim=0 0 0 0]{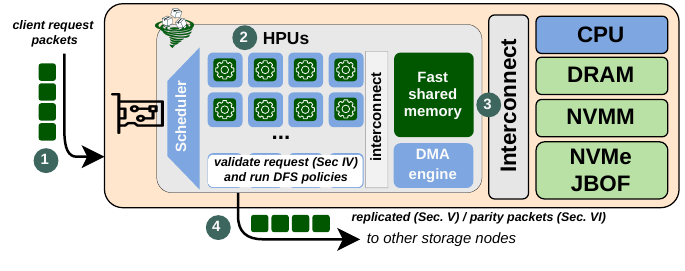}
	\vspace{-2.6em}
	\caption{Network-offloaded DFS policies overview.}
	\vspace{-1em}
	\label{fig:spin_dfs_overview}
\end{figure}

\section{NIC-offloaded DFS policies}

Figure~\ref{fig:spin_dfs_overview} shows an overview of our approach to network-offloaded DFS policies. Clients issue data access requests to storage nodes~\circlegreen{1}, triggering DFS policies that are executed on the HPUs of the sPIN-enabled NIC~\circlegreen{2}. 
The DFS handlers authenticate requests coming from the client (see Section~\ref{sec:auth}), write data to host~\circlegreen{3}, and enforce other DFS policies~\circlegreen{4} such as replication (Section~\ref{sec:replication}) and erasure coding (Section~\ref{sec:ec}).
By running DFS policies as sPIN handlers, we can perform actions (e.g., forward packets to the next replica node) before data reaches host memory via the system interconnect, saving latency for small writes (e.g., a PCIe round-trip can take up to 400 ns~\mbox{\cite{kalia2020challenges}}), and leveraging packet-level parallelism for larger ones (i.e., composed of multiple network packets).

We do not focus on a specific storage medium. Instead, we focus on showing the benefits of offloading DFS policy enforcement to the NIC and assume that the storage medium can digest data at network bandwidth or higher. For example, with in-memory or non-volatile-main-memory (NVMM) based DFSs, handlers would write directly to main memory, as any other RDMA DFS~\cite{yang2020filemr, yang2019orion}. For DFSs targeting NVMe just-a-bunch-of-flash (JBOF), handlers would directly issue NVMe writes via the system interconnect (e.g., PCIe).

\subsection{Client request format}\label{sec:wire_format}

\begin{figure}[h]
	\centering
	\vspace{-1em}
	\includegraphics[width=0.85\columnwidth, trim=0 0 5 0]{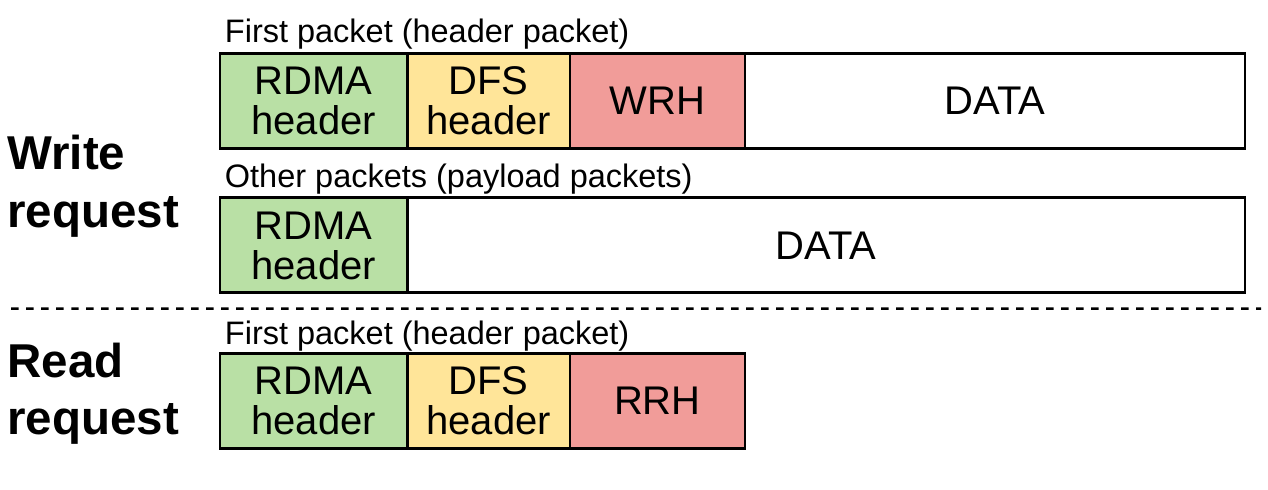}
	\vspace{-1.2em}
	\caption{Packet format for write and read requests.}
	\vspace{-1em}
	\label{fig:rw_packets}
\end{figure}

Figure~\ref{fig:rw_packets} shows the layout of write and read requests.
A write request consists of: an RDMA header (e.g., InfiniBand or RoCEv2); a generic DFS header carrying information to identify the request (e.g., operation type) and to authenticate it; a write request header (WRH) that carries write-specific information, such as the replication strategy and the number and coordinates of the replica nodes and respective address where data must be replicated. 
The headers are followed by the packet payload (i.e., the data to write).
If the write spans multiple network packets, then only the first packet carries DFS-specific headers, while others consist of the RDMA header and the continuation of the data to write. 
A read request carries the RDMA header, the DFS header, and a read request header (RRH) with read-specific information.

We assume that request headers (DFS and WRH/RRH) always fit in a single network packet. 
This assumption is realistic for, e.g., RoCE networks, where packet sizes are limited by Ethernet maximum transmission unit (MTU), which typically ranges between \mbox{1.5 KiB} and \mbox{9 KiB} (jumbo frames).

\subsection{sPIN handlers}

Listing~\ref{lst:handlers} shows the pseudocode of generic sPIN handlers for offloading DFS components. Each handler takes two arguments: a task descriptor and the pointer to a network packet. The task descriptor contains information about the corresponding execution context, such as NIC memory allocated for and shared by the handlers running on packets matched by this specific execution context.

There are two types of DFS-specific tasks that can be executed on each request: per-request tasks and per-packet tasks. 
Per-request tasks can be expressed in \texttt{DFS\_request\_init} and \texttt{DFS\_request\_fini}, which are called once per message, i.e., in the header and completion handlers, respectively. Per-request tasks can be used for, e.g., validating the request before processing other packets of the same request (i.e., sPIN guarantees that the payload handlers of a message are executed after the header handler of that message completes).
As the completion handler is executed only once all packets of a message have been processed, the DFS can use \texttt{DFS\_request\_fini} for finalizing the handling of the request: e.g., send the write acknowledgment back to the client. Additionally, a payload handler is executed for each packet (including header and completion). Here, the DFS can run per-packet actions such as copying the payload to the storage media and sending the packet to the next replica node.

These handlers are triggered for all incoming client requests: i.e., the first packet of any request triggers a header handler, the last triggers a completion handler, and all packets trigger a payload handler.
They are persistent and do not need to be installed on a per-request basis nor require an established connection between clients and storage nodes.

\begin{lstlisting}[style=customc, caption=Generic sPIN handlers for offloading DFS tasks., label=lst:handlers]
	void header_handler(spin_task_t* task, pkt_t* pkt) {
		dfs_state_t* state = (dfs_state_t*) task->mem;
		
		bool accept_next_pkts = DFS_request_init(state, pkt);
		// DFS_request_init sends NACK if request auth fails
		
		int req_idx = task->flow_id;
		state->req_table[req_idx].greq_id = pkt->dfs.greq_id;
		state->req_table[req_idx].accept = accept_next_pkts;
	}
	
	void payload_handler(spin_task_t* task, pkt_t* pkt) {
		dfs_state_t* state = (dfs_state_t*) task->mem;
		int req_idx = task->flow_id;
		
		if (state->req_table[req_idx].accept) 
		DFS_request_process_pkt(state, pkt);
		//else packet is dropped
	}
	
	void tail_handler(spin_task_t* task, pkt_t* pkt) {
		dfs_state_t* state = (dfs_state_t*) task->mem;
		int req_idx = task->flow_id;
		
		if (state->req_table[req_idx].accept) 
		DFS_request_fini(state, pkt);
		//else packet is dropped
	}
\end{lstlisting}

\subsubsection{Data persistence}
A write operation completes when the data reaches the storage target. If client requests are handled on CPUs of the storage nodes, the DFS can wait for the data to be written to the memory target before acknowledging the client. In RDMA-based DFSs, this task is slightly more complex. There, when a client gets an RDMA write completion event, the data could be on the storage node persistent memory or still be buffered somewhere between the NIC and the storage target (e.g., PCIe buffers). To overcome this limitation, the client can issue an RDMA read immediately after a write to flush DMA buffers, triggering a read-after-write dependency~\cite{kalia2020challenges}. 

While RDMA extensions are being proposed to introduce an RDMA \emph{flush} operation~\cite{talpey2016rdma} that could be merged together with a write to save the additional RDMA read latency, this is a good example of the advantages of using SmartNICs.
With sPIN, packet handlers can explicitly issue writes to the storage medium, making sure that the data is flushed before acknowledging the client, as it would happen on the CPU. Additionally, a sPIN handler can overlap other activities (e.g., running other DFSs policies, such as replication) while waiting for the data to be flushed to the storage target.

\subsubsection{Scalability}
Each write request requires to keep a state in the NIC for the whole operation duration (i.e., \texttt{req\_table} entries in Listing~\ref{lst:handlers}). Each entry is a write descriptor that takes 77 bytes and stores the current status of the request plus information that is carried only by the header packet and that is needed by payload handlers. For example, for data replication (see Section~\ref{sec:replication}), we use a source-based approach where the WRH carries replica addresses. As we need to forward all packets of the write request, we store these addresses in the per-request state to make forwarding possible. 

In PsPIN, each one of the four compute clusters is equipped with a 1 MiB single-cycle access memory (L1). Additionally, there is an off-cluster NIC memory of 4 MiB (L2). We store client request descriptor into L1 and use L2 as a swap-out area. In total, we have 6 MiB of available memory to store client request entries, while the remaining 2 MiB are used to store DFS-wide state. This allows us to serve up to $\sim$82~K concurrent writes for each storage node. If a client request cannot be served because of lack of space, the request is denied, and the client will retry later.
\begin{figure}[h]
	\centering
	\vspace{-1em}
	\includegraphics[width=\columnwidth, trim=0 0 0 0]{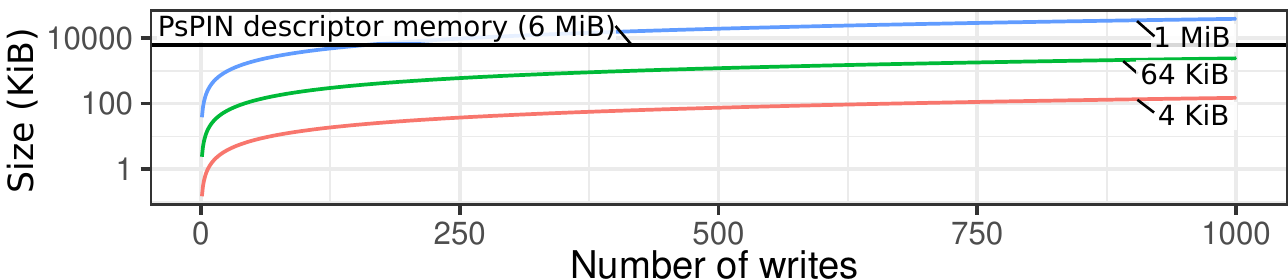}
	\vspace{-2em}
	\caption{Worst-case required NIC memory versus number of writes and write sizes. The horizontal line indicates the amount of NIC memory required.}
	\vspace{-0.5em}
	\label{fig:state_scal}
\end{figure}

The number of writes served by a storage node at any given time depends on the write size, sPIN handler running times, network bandwidth, and network state (e.g., congestion). We apply Little's law to make a worst-case analysis of the average number of writes being served at any given time by a storage node, assuming a constant flow of fixed-size writes arrive at full bandwidth. Figure~\ref{fig:state_scal} shows the required NIC memory to handle a specific number of writes (x-axis) and for different write sizes. In this analysis, we assume that sPIN handlers are not a bottleneck. Detailed data about sPIN handler running times are reported in the following sections.

\subsection{Full-system design considerations}
To offload policies to the NIC, the DFS software running on the storage nodes CPU instantiates a PsPIN execution context made of DFS handlers and a NIC memory region storing the DFS state.
The execution context is installed into the storage node NIC and matches all incoming RDMA packets. The DFS software running on the CPU can communicate with sPIN handlers by writing to NIC memory (e.g., to update encryption keys, see Section~\mbox{\ref{sec:auth}}). On the other side, handlers can communicate with the DFS software through application-specific event queues (e.g., error conditions, logging information, and other policy-specific events).

The DFS endpoint is similar to RDMA-based DFS. A client must be able to retrieve metadata to build read/write requests (e.g., addressing information for primary and secondary storage nodes). 
On the storage nodes, requests can be handled either by PsPIN, as we show in this work, or by the DFS software running on the storage node CPU (e.g., by appending requests to RPC command queues via RDMA~\mbox{\cite{kalia2016fasst, taranov2021corm}}). For example, the execution context can be configured to steer requests to host memory, bypassing PsPIN, if the SmartNIC is not keeping up with line rate (e.g., overwhelmed by writes requiring erasure coding, see Section~\mbox{\ref{sec:ec}}).

\subsection{Experimental methodology}
Until now, we described sPIN, the programming model that best fits the principles of Section~\ref{sec:dfs_intro}, and discussed the skeleton of the sPIN handlers used to offload the DFS policies. The next sections show how three different DFS policies can be offloaded to the NIC and the respective performance benefits.

The evaluation shown in the following sections is performed through cycle-accurate and functional simulations. We use the PsPIN toolchain to produce cycle-accurate timings of sPIN handlers and the Structural Simulation Toolkit (SST)~\cite{rodrigues2011structural} to simulate multi-node scenarios (e.g., clients and storage nodes). PsPIN handlers are compiled with a PULP-custom version of GCC 7.1.1 (riscv32, \texttt{-O3 -flto}).
We configure SST to simulate a 400 Gbit/s network, with a maximum transmission unit (MTU) of 2048 B and 20 ns link latency.

\section{Client request authentication}
\label{sec:auth}

To access data, clients first get metadata information and then directly contact the storage nodes to access data. In this scenario, storage nodes can either trust or authenticate client requests. However, it is not always possible to fully trust clients in shared distributed systems, as a malfunctioning or malicious application/node could bring the DFS to an inconsistent state. 

\begin{figure}[t]
	\centering
	\vspace{-0.4em}
	\includegraphics[width=\columnwidth, trim=0 0 0 0]{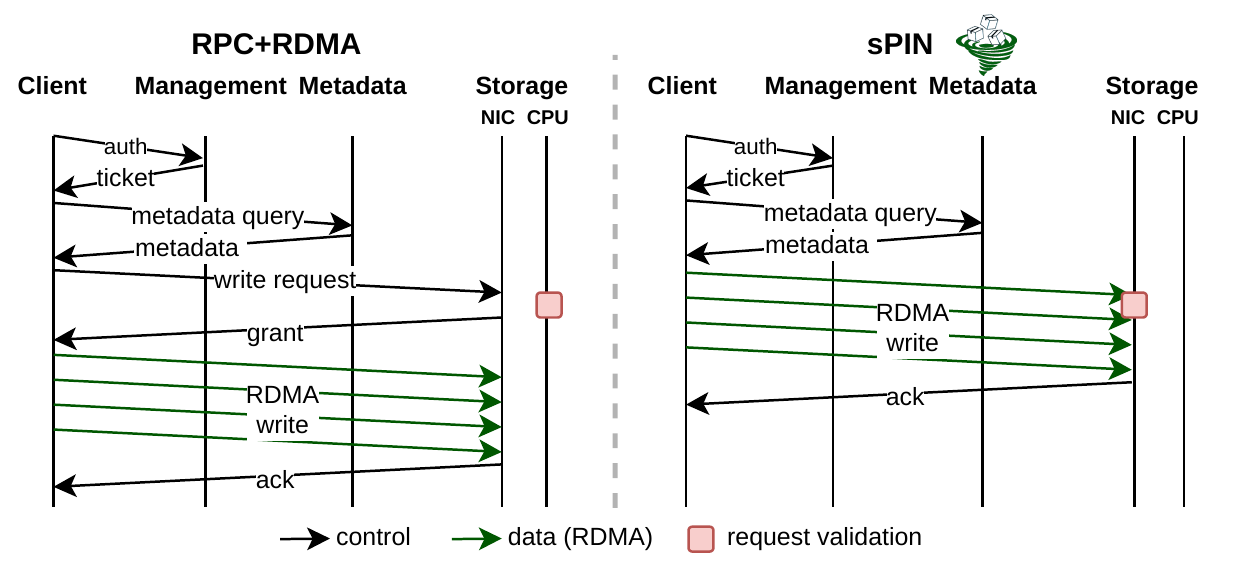}
	\vspace{-2.2em}
	\caption{Authentication strategies. Left: authentication performed on the host CPU (RPC+RDMA). Right: on the NIC with sPIN. The storage node is represented with both NIC and CPU to show where activities are executed.}
	\vspace{-1em}
	\label{fig:one_sided_access}
\end{figure}

The way client requests are authenticated is DFS-specific. While RDMA provides protection via \emph{rkeys} (i.e., a node can access a remote region only if it has the respective rkey), this offers limited authentication capabilities for client requests. For example, having a single \emph{rkey} for all files would not prevent clients from accessing data they do not own, while having a rkey for each file would require registering $\#files \times \#rights$ memory regions, which can lead to scalability issues~\cite{wang2021star}. 
On the other side, validating requests on the CPU of the storage nodes increases overhead, requiring either to buffer the full write before committing it to the storage target (losing the zero-copy benefit of RDMA) or to introduce an additional network round-trip before the RDMA-write. Figure~\ref{fig:one_sided_access} (left) sketches the scenario where CPU-based request authentication is performed. In this case, after the client authenticated itself with the management node and queried the metadata node, it sends a request to the storage node. The request is validated on the CPU of the storage node and, after that, the actual RDMA data transfer can begin.

The sPIN case is shown in Figure~\ref{fig:one_sided_access} (right). The validation is offloaded to the NIC and implemented in the \texttt{DFS\_request\_init} function of Listing~\ref{lst:handlers}. This allows the client to directly issue the RDMA write as the request validation will be performed on-the-fly by sPIN handlers, saving the extra RTT required for authenticating the client. 

The way requests are validated depends on the threat model: 
if we trust both the clients and network (e.g., sRDMA~\cite{taranov-srdma}), then the ticket can be a plain-text secret given to the client by the management node and checked by the handlers. Nobody can read the ticket from the client or intercept it in the network. 
DFSs like Orion~\cite{yang2019orion}, where storage nodes are accessed by the clients via RDMA, assume this threat model. 

If clients are not trusted but the network is, then we need authentication capabilities: the client gets a ticket from the metadata node that contains a capability descriptor. This descriptor determines the operations that the client is allowed to perform and the data it can access. 
The handlers verify the capability, which is signed with a key shared among DFS services, and check that the requested operation is allowed by the capability~\cite{gobioff1997security}. In this work, we assume this threat model.
If the network is not trusted, then handlers need to authenticate each network packet in order to exclude tampering.

We analyze the impact of processing packets through sPIN on the overall write latency. The write latency is the time spanning from issuing the write request to receiving the respective write response. 
In this case, we consider writes where only client request authentication is performed, without additional DFS policies (e.g., replication or erasure coding).
We consider the following write protocols:
\begin{itemize}[noitemsep,topsep=1pt,parsep=0pt,partopsep=0pt,leftmargin=10pt]

	\item \textbf{RPC+RDMA}: The client first sends the write request to the storage node via RPC. The RPC handler, executed on the CPU of the storage node, runs DFS policies (e.g., validating the client request) and issues an RDMA read towards the client for getting the data to store.
	\item \textbf{RPC}: The client sends the write request and the data to store to the storage node via RPC. The storage node buffers the data to write, runs DFS policies, and eventually stores the data in the storage target.
	\item  \textbf{sPIN}: The client sends the write request and data to store to the storage node in a single RDMA write. Packets are intercepted by sPIN at the NIC of the storage node, where client requests are authenticated by sPIN handlers.
	\item \textbf{Raw writes:} this is our speed-of-light scenario. No DFS policies are enforced on incoming writes. The client issues a single RDMA write to the storage node. 
\end{itemize}

\subsection{Request authentication overhead}\label{sec:ev:simple_writes}
\begin{figure}[t]
	\centering
	\subfigure{{\includegraphics[width=0.9\columnwidth]{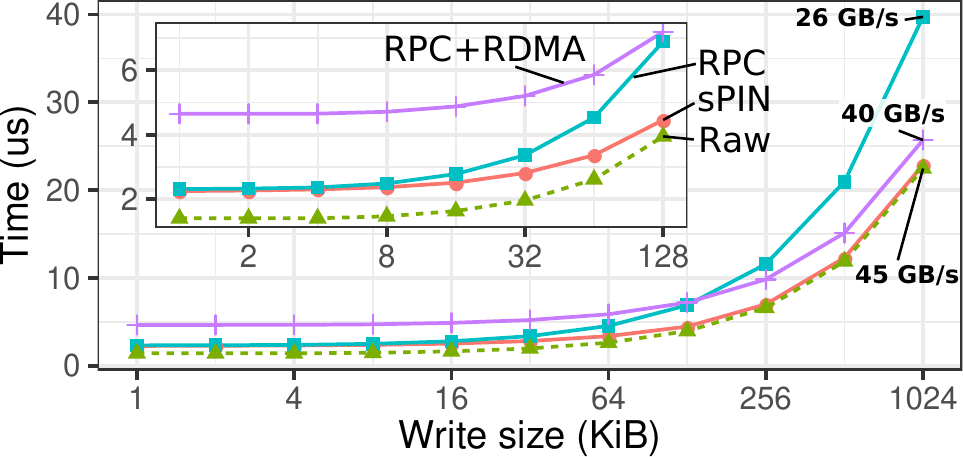} }}%
	\vspace{-0.5em}
	\caption{Write latency with different protocols and write sizes. Raw writes are reported as a (speed-of-light) reference as they do not enforce any policy.}%
	\vspace{-1em}
	\label{fig:data:simple_writes}%
\end{figure}

\begin{figure}[h]
	\vspace{-1.5em}
	\includegraphics[width=\columnwidth]{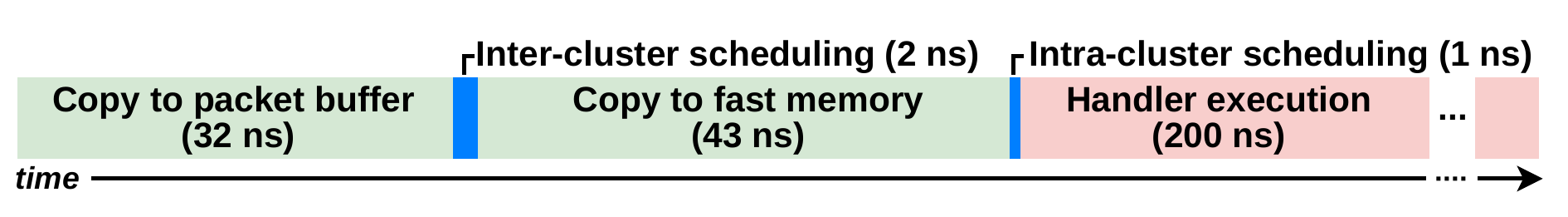}
	\vspace{-2em}
	\caption{Packet processing overheads in PsPIN (for 2 KiB packets). 
	}%
	\vspace{-1em}
	\label{fig:simple_write_costs}%
\end{figure}

Figure~\ref{fig:data:simple_writes} shows write latencies for different write sizes. The sPIN handlers validate client requests by checking the capability carried in the write request header (see Section~\ref{sec:wire_format}). For RPC and RCP+RDMA, the same validation is performed on the CPU. For large writes, RPC is penalized by the additional memory copy needed to buffer the write while the request is validated. We notice how sPIN introduces small overheads over raw writes, which do not perform validation of incoming write requests. For small writes, sPIN pays the latency of having packets traverse the on-NIC accelerator and validate requests, hence it shows higher overheads (up to 27\%) than the raw writes. Figure~\ref{fig:simple_write_costs} breaks down the overheads of processing packets in PsPIN: a packet is first copied into the packet buffer (32 cycles for a 2 KiB packet), then scheduled to one of the four processing clusters (2 cycles). At this point, the packet is copied into the cluster-local, single-cycle access memory (43 cycles) and finally scheduled to an idle HPU (1 ns). The DFS handler that validates client requests takes 200 cycles. For large writes, the per-request validation performed only on the first packet of the write becomes negligible, making sPIN-processed writes approach the RDMA (speed-of-light) latency.

\begin{figure*}[t]
	\centering
	\includegraphics[width=\textwidth, trim=0 0 0 0]{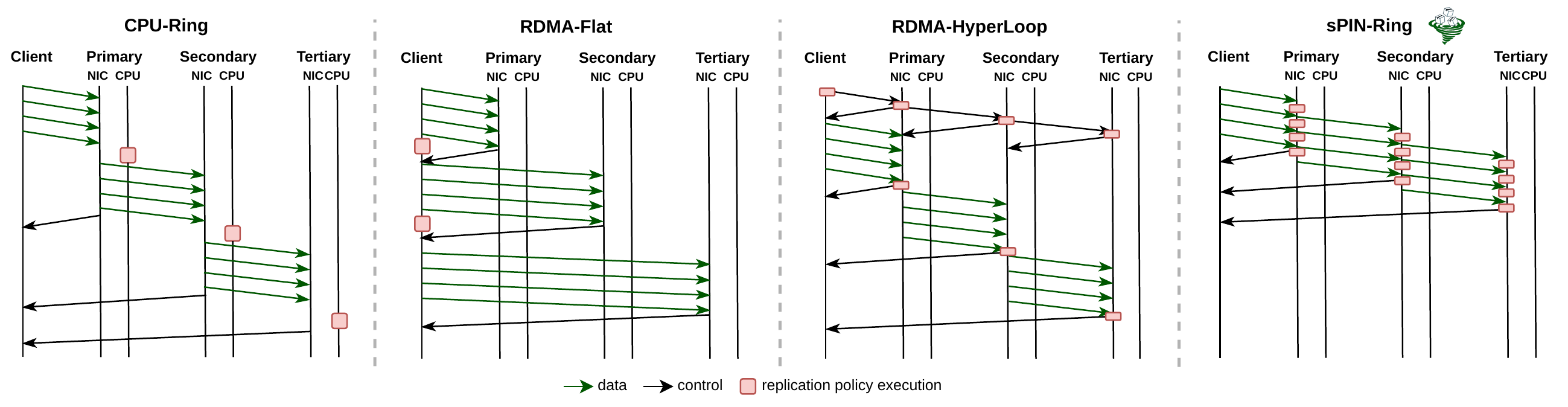}
	\vspace{-1.5em}
	\caption{Data replication policy implemented with different strategies. CPU-Ring involves the CPU of the storage nodes to implement a ring broadcast. With RDMA-Flat, clients write directly to the storage node memory but need to make $k$ different RDMA writes, one for each replica. In the RDMA-Hyperloop case, the RDMA ring needs to be configured with a smaller, pre-defined broadcast for metadata (i.e., for updating RDMA work queue entries to serve the actual data broadcast). With sPIN, the policy is offloaded to the NIC of the storage nodes, that propagate the data on a per-packet basis.
	}
	\vspace{-1em}
	\label{fig:replication_diagram}
\end{figure*}

\section{Data replication}
\label{sec:replication}

This DFS policy replicates data on $k$ storage nodes, where $k$ can be either a global, per-pool, or per-file parameter.
In this way, data can survive the failure of $k-1$ storage nodes. To enforce replication, the data written by a client should be propagated to $k$ storage nodes. 
In the following, we use $k$ to indicate the replication factor, that is the number of nodes on which data is replicated.

Figure~\ref{fig:replication_diagram} shows different strategies for data replication. 
If the CPU of the storage nodes is involved, then data can be broadcasted among the $k$ storage nodes following different strategies (i.e., broadcast schedules). The optimal broadcast schedule depends on $k$, the data size, and the interconnection network characteristics~\cite{karp1993optimal, alexandrov1995loggp}. 

In RDMA-based DFSs, data replication can be performed by either delegating the replication process to the client (RDMA-Flat, i.e., the client performs $k$ different RDMA writes) or with pre-posted RDMA operations that need to be configured by the client out-of-band~\cite{kim2018hyperloop} (RDMA-HyperLoop). 

With sPIN, storage nodes can exploit the exposed NIC computing capabilities to distribute replicated data using different broadcasting strategies without involving the host CPU. Specifically, we consider two different replication strategies: \textit{ring}, where each replica sends to another replica only, and \textit{binary tree}, where each replica sends to two children. We show in Figure~\ref{fig:replication_diagram} an example where the sPIN handlers on the storage node propagate the data in a pipeline with the replica nodes virtually arranged on a ring (sPIN-Ring). Since each packet must be forwarded, the broadcast algorithm is implemented in the \texttt{DFS\_request\_process\_pkt} function of Listing~\ref{lst:handlers}.

\subsection{Broadcast schedules in sPIN}

Any broadcast schedule offloaded with sPIN is naturally pipelined on network packets.
In particular, for participating in the broadcast, each packet handler needs to (1) identify their position in the broadcast tree (a \textit{ring} can be seen as a unary tree) and (2) send a copy of the data to its children if any.

We require that the write request header (see Figure~\ref{fig:rw_packets}) carries the following information:
\begin{itemize}[noitemsep,topsep=1pt,parsep=0pt,partopsep=0pt,leftmargin=10pt]
	\item Replication strategy: \textit{ring} or \textit{pipelined binary tree} (pbt);
	\item Virtual rank: an ID that identifies the node in the tree;
	\item Replica coordinates: a list of tuples representing the replica nodes. Each tuple includes the network address of the replica and the respective storage address.
\end{itemize}
This information is contained in the first packet of the write request and needs to be also propagated to the handlers processing the subsequent packets. For this reason, we keep an array of replica coordinates (\emph{coord\_array}) in the DFS state stored in NIC memory:
this array has a length equal to the arity of the broadcast tree (i.e., $1$ if \emph{ring}, 2 if \emph{pbt}) and is filled in by the header handler. The header handler uses the virtual rank, the replication strategy, and the list of replica coordinates to identify the children where to send data. The payload handlers check the DFS state and send the data to all replica coordinates into \emph{coord\_array}.

By having the request carrying information on how to progress the communication, we create a client-driven broadcast that does not require the involved storage nodes to keep CPU-initialized stateful data structures to progress it (e.g., to know where to forward the data). We still need a stateful data structure (i.e., the \emph{coord\_array}), but that can be initialized when the first packet of the request arrives.

\begin{figure*}[t]
	\subfigure{{\includegraphics[width=0.38\textwidth]{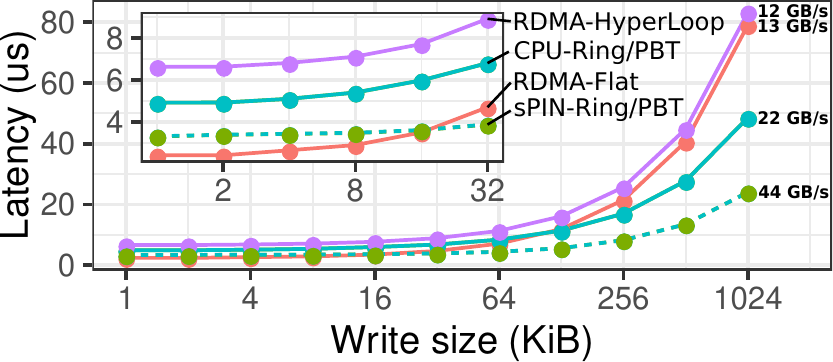} }}%
	\subfigure{{\includegraphics[width=0.38\textwidth]{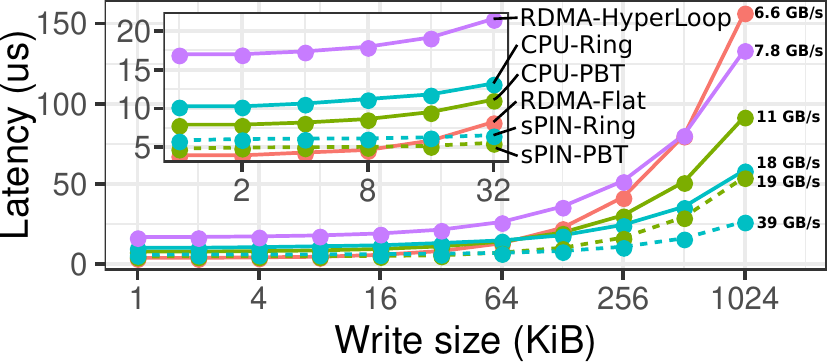} }}%
	\subfigure{{\includegraphics[width=0.22\textwidth]{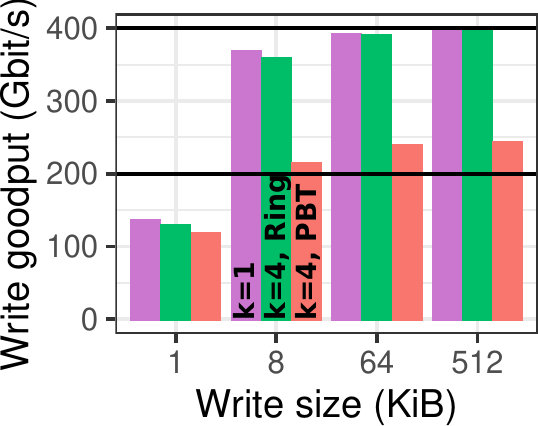} }}%
	\vspace{-0.5em}
	\caption{Left: Write latency with replication factor ($k$) set to 2 for different write sizes and replication strategies. Center: Write latency with replication factor ($k$) set to 4. Right: Goodput sustained by a single storage node for different write sizes and offloaded replication strategies.}%
	\label{fig:data:rep:latency}%
	\vspace{-0.5em}
\end{figure*}

\subsection{Data replication performance}

To analyze how per-packet processing impacts data replication performance, we measure the write latency for different data sizes and replication factors. We consider different replication strategies, ranging from fully offloaded with sPIN to CPU based:
\begin{itemize}[noitemsep,topsep=1pt,parsep=0pt,partopsep=0pt,leftmargin=10pt]	
	\item \textbf{CPU-Ring.} Replica nodes are virtually arranged in a ring. The CPU of the storage nodes is notified of incoming writes and forwards data to the next node in the (pipelined) ring.
	\item \textbf{CPU-PBT.} Similar to \emph{CPU-Ring} but replica nodes are arranged in a binary tree.
	\item \textbf{RDMA-Flat.} A client issues as many writes as the number of replica nodes.
	\item \textbf{RDMA-HyperLoop.} A client first updates the work-queue-elements (WQEs) on the RDMA NICs of the storage nodes and then starts an RDMA ring broadcast~\cite{kim2018hyperloop}. The WQEs need to be updated in order to define where to write the data on the storage node. 
	\item \textbf{sPIN-Ring}. Replicas are arranged in a ring. Data is forwarded by sPIN handlers running on the NIC of the storage nodes. Data is naturally pipelined on network packets.
	\item \textbf{sPIN-PBT}. Similar to \emph{sPIN-Ring} but replica nodes are arranged in a binary tree.
\end{itemize}
RDMA-Flat and RDMA-HyperLoop do not enforce request validation and fully trust clients. We report data from pipelined executions with optimal chunk size for all non-sPIN strategies implementing a ring and binary-tree broadcast.

\subsubsection{Write latency}

Figure~\mbox{\ref{fig:data:rep:latency}} (left) and Figure~\mbox{\ref{fig:data:rep:latency}} (center) show latencies of writes with replication factors of $k$=2 and $k$=4, respectively. Each plot shows the write latency as function of the write size and for different replication strategies.
With $k$=2, there are no differences between \mbox{ring} and \mbox{pbt} replication strategies as the primary storage node has only one children. In both scenarios, RDMA-Flat provides the lowest latency for small writes (up to 16 KiB), being up to 27\% faster than sPIN-Ring for 1 KiB writes and $k$=2. However, RDMA-Flat would require an additional round-trip per replica to validate write requests. For writes bigger than 16 KiB, the data injection cost paid by the client starts impacting RDMA-Flat performance, making sPIN-based solutions faster.
RDMA-Hyperloop is penalized by configuration overheads (i.e., writing of WQEs at storage nodes, see Figure~\mbox{\ref{fig:replication_diagram}}). These overheads get better amortized for long replication chains ($k$=4) and large message sizes.
Overall, sPIN-based solutions achieve up to 2x and 2.16x lower latency (w.r.t. the best alternative) for $k$=2 and $k$=4, respectively.
CPU-based replication strategies are negatively impacted by the cost of moving data to and from host memory.

\begin{figure}[b]
	\vspace{-2em}
	\subfigure{{\includegraphics[width=0.5\columnwidth]{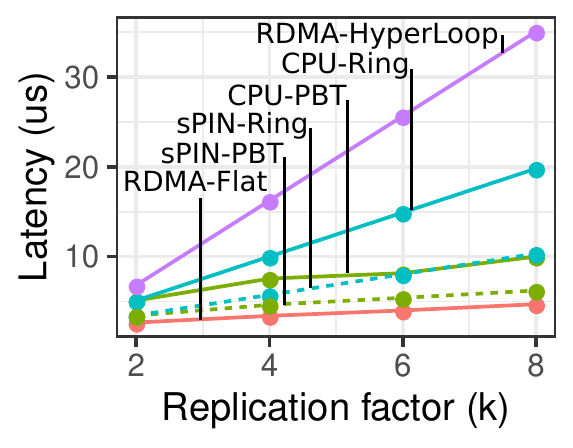} }}%
	\subfigure{{\includegraphics[width=0.5\columnwidth]{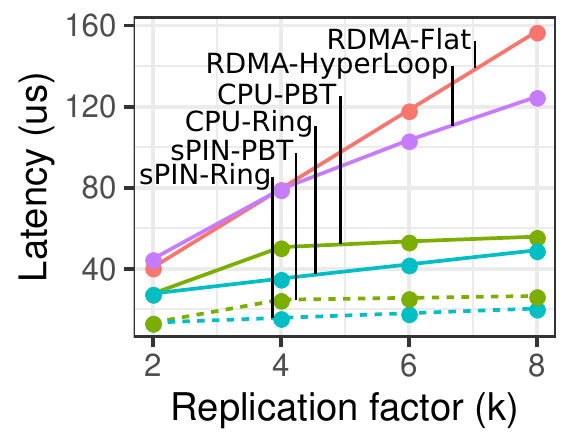} }}%
	\vspace{-1.2em}
	\caption{Write latency for small (4 KiB, left) and large (512 KiB, right) writes with different replication strategies and replication factors ($k$).}%
	\label{fig:data:rep:details}%
\end{figure}

\subsubsection{Write goodput}
Figure~\ref{fig:data:rep:latency} (right) shows the goodput sustained by a single network-accelerated storage node for different write sizes and replication strategies.  This is the amount of data per second, excluding headers, that can be ingested by a storage node.

A 1 KiB write fits in a single packet (MTU is 2 KiB) that triggers all handlers (header, payload, and completion). Since each packet triggers three handlers, the overall throughput that sPIN can sustain is limited. With larger writes, the number of packets per write increases, better amortizing header and completion handlers costs (executed once per write). Starting from 8 KiB writes sPIN-Ring line rate.  
Writes replicated with sPIN-PBT achieve about half the bandwidth because, for each incoming packet, two new packets (i.e., one per children) must be sent out. However, the bandwidth cost is compensated by a lower overall latency of the binary tree for small writes and/or large values of $k$ (see Figure~\ref{fig:data:rep:details} (left)), allowing to achieve write latencies better or similar to sPIN-Ring in those cases.

\subsubsection{Varying the replication factor} 
Figure~\mbox{\ref{fig:data:rep:details}} shows the write latency as function of the replication factor for 4 KiB (left plot) and 512 KiB (right plot) writes. 
For small writes, \mbox{RDMA-Flat} has the lowest latency for any replication factor. 
For large writes, the injection cost increases, limiting the performance of \mbox{RDMA-Flat} that grows linearly with $k$. 
CPU-based pipelined strategies become more efficient for large replication factors, but they are still penalized by memory-copy overheads.
On the other hand, sPIN-based versions are less sensible to $k$ because of the smaller per-packet overheads. As expected, \mbox{\emph{pbt}}-based replication performs better than \mbox{\emph{ring}}-based ones for small writes and large values of $k$. 

\begin{figure}[h]
	\vspace{-0.5em}
	\centering
	\includegraphics[width=1\columnwidth]{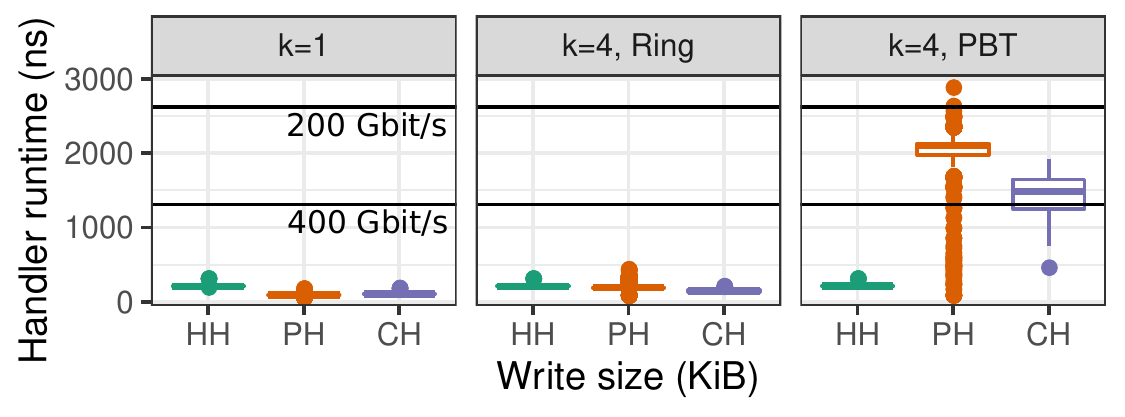}%
	\vspace{-1em}
	\caption{Handlers running time for writes with replication. Left: no replication; Middle: 4 replicas, sPIN-Ring; Right: 4 replicas, sPIN-PBT.}%
	\vspace{-0.5em}
	\label{fig:data:rep:handlers}%
\end{figure}

\subsubsection{Handlers runtime analysis}

Figure~\ref{fig:data:rep:handlers} shows handler running times for different handler types and replication strategies. We plot lines indicating the cycle budget available to each handler to sustain 400 Gbit/s and 200 Gbit/s line rates with 2 KiB packets. We observe that the duration of handlers for writes without replication ($k=1$) and writes with ring-based replication ({sPIN-Ring}) always stays below the 400 Gbit/s budget. 
The higher running times of {sPIN-PBT} handlers justify the lower write goodput discussed above. 
As shown in Table~\ref{table:rep-handlers}, the longer duration of these handlers does not correspond to an higher complexity but to a lower number of instructions per cycle (IPC). This is a consequence of the fact that the egress network bandwidth is limited (400 Gbit/s) and that each incoming packet generates two outgoing ones.

\begin{table}[h]
	\vspace{-0em}
	\scriptsize
	\setlength{\tabcolsep}{1.5pt}
	\begin{tabularx}{\columnwidth}{p{1.5cm}XXXXXXXXX}
		\multirow{2}{*}{\textbf{Type}}  & \multicolumn{3}{c}{\textbf{Duration (ns)}} & \multicolumn{3}{c}{\textbf{Instructions}} & \multicolumn{3}{c}{\textbf{IPC}} \\
		& HH & PH & CH                               & HH & PH & CH                              & HH & PH & CH                     \\
		\midrule
		\textbf{k=1} 		& 211 & 92    &  107 &    120 & 55 & 66     & 0.57 & 0.60 & 0.62 \\
		\textbf{k=4, Ring} 	& 212 & 193   &  146 &    120 & 105 & 65    & 0.57 & 0.54 & 0.44 \\
		\textbf{k=4, PBT} 	& 214 & 2106  & 1487 &    120 & 130 & 82    & 0.56 & 0.06 & 0.06 \\
		\bottomrule
	\end{tabularx}
	\vspace{0.3em}
	\caption{Handler statistics for different replication strategies. 
	}
	\vspace{-1em}
	\label{table:rep-handlers}
\end{table}

\section{Erasure coding}
\label{sec:ec}

The main disadvantage of replication is the storage cost, which is linear in the replication factor.
With erasure coding (EC), data is split into $k$ chunks and stored together with $m$ parity chunks. The $k+m$ chunks are normally stored on different storage nodes and failure domains. In case of failure, the missing chunks can be recovered by using the remaining ones.
Reed-Solomon~\cite{reed1960polynomial} codes (RS) are erasure codes employed in a variety of storage systems. 
RS is maximum distance separable, meaning that a $RS(k, m)$ code can survive up to $m$ corrupt chunks. Also, RS codes are \emph{systematic}: $k$ of $k+m$ encoded chunks are identical to the original $k$ data chunks and can be read without any decoding process.

\begin{figure}[h]
	\centering
	\vspace{-1em}
	\includegraphics[width=0.7\columnwidth, trim=0 0 5 0]{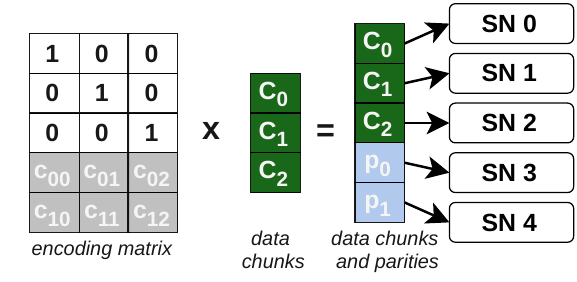}
	\vspace{-1em}
	\caption{Encoding an object with $RS(3,2)$. SN: Storage Node.}
	\vspace{-1em}
	\label{fig:ec_layout}
\end{figure}

Figure~\ref{fig:ec_layout} shows an example for $RS(3, 2)$: the encoding matrix ($5 \times 3$) is multiplied by the data chunks ($3 \times 1$), obtaining a $5 \times 1$ matrix with the $3$ data chunks and $2$ parity chunks.

\subsection{INEC-TriEC}
Offloaded EC schemes have been investigated by Shi et al.~\cite{shi2019triec, shi2020inec}. In particular, TriEC is a distributed EC scheme where encoding/decoding activities are distributed among multiple storage nodes.
With INEC~\cite{shi2020inec}, Shi et al. introduce a set of in-network EC primitives to accelerate both encoding and decoding phases of different EC schemes, including TriEC (here defined as INEC-TriEC). 
TriEC distributes the encoding of data chunks to different storage nodes, generating streams of data to be processed (i.e., data chunks). As sPIN enables application-defined data stream processing on the NIC, it is a good
candidate for accelerating TriEC.

\begin{figure}[t]
	\centering
	\vspace{-1em}
	\includegraphics[width=\columnwidth, trim=0 0 0 0]{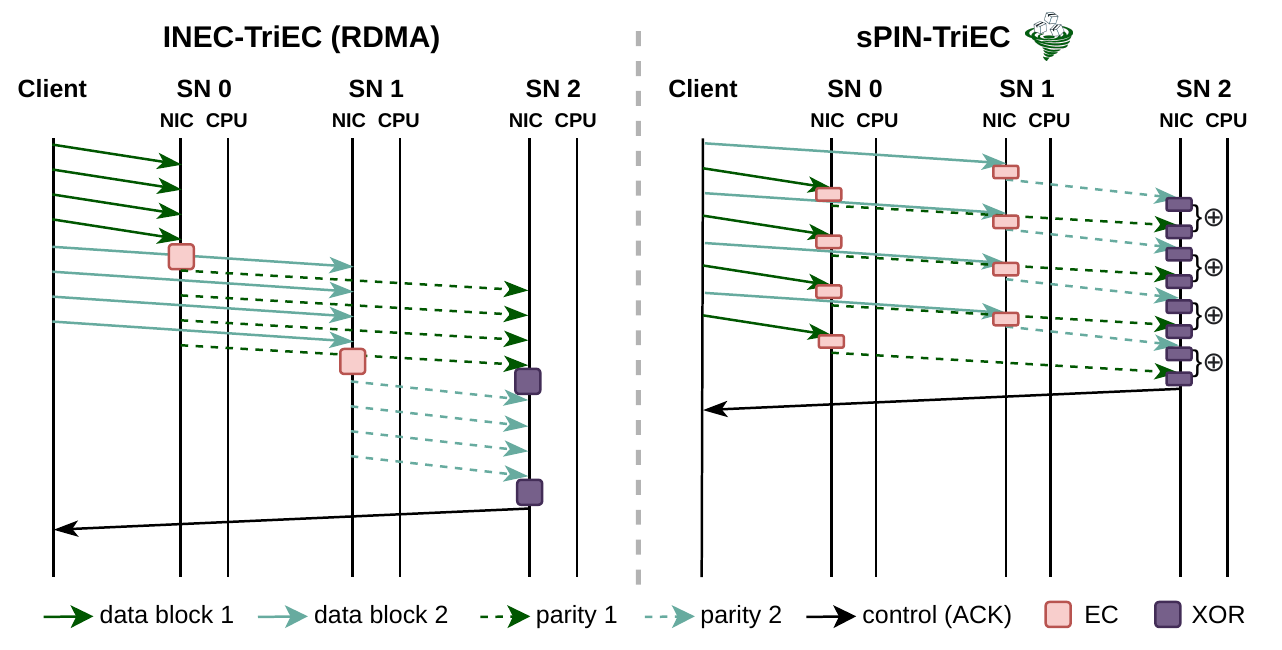}
	\vspace{-2em}
	\caption{Diagram of RS(2,1) with INEC-TriEC (left) and sPIN-TriEC (right).}
	\vspace{-1.6em}
	\label{fig:ec_diagram}
\end{figure}

Figure~\ref{fig:ec_diagram} (left) shows the erasure coding of a data block in an $RS(2,1)$ scheme with INEC-TriEC. The client sends two different data chunks to two different storage nodes (\emph{SN 0} and \emph{SN 1}), where data is moved to the main memory, as for normal RDMA write (not shown in the figure). Once the transfer is complete, the EC computation is triggered and executed directly on the NIC, reading data from main memory and computing the parity chunks. These are sent to a different storage node (\emph{SN 2} in the example). In general, for each data chunk, $m$ different intermediate parity chunks are generated and distributed to different parity nodes.

\subsection{sPIN-TriEC} \label{sec:ec_handlers}

With sPIN, the TriEC approach can be re-interpreted with a per-packet vision, allowing to encode data in a streaming fashion, hence avoiding waiting for the full chunk of the data to be received (and copied into host memory) first.
We focus on the encoding part since its latency contributes to the write latency if strong consistency is required. The decoding process should preferably be performed offline to not impact write latency. For example, monitoring services can check the status of the storage nodes and start the recovery process if some of them become unreachable~\cite{borthakur2008hdfs, weil2006ceph}. 

We consider a write request format similar to the one discussed for the replication strategies (see Section~\ref{sec:replication}). We assume a write request header carrying enough information to allow the storage node to identify its role in the distributed algorithm. In particular, the write request header carries:
\begin{itemize}[noitemsep,topsep=1pt,parsep=0pt,partopsep=0pt,leftmargin=10pt]
	\item Erasure coding scheme: \emph{RS-k-m}, where $k$ and $m$ are the number of data and parity chunks, respectively;
	\item Role: indication of whether this node stores data or parity chunks, determining the actions performed by the handlers;
	\item Parity node coordinates: the coordinates of the parity nodes. 
\end{itemize}
Replication and erasure coding are normally mutually exclusive: a file is either replicated or erasure-coded. For this reason, the write request header carries a \emph{resiliency strategy} option, telling us whether
replication, EC, or no resiliency schemes should be used for this write. This option is followed by either replication or EC parameters.

\subsubsection{Sending packets} In Figure~\ref{fig:ec_diagram} (right), the client transmits packets to \emph{SN 0} and \emph{SN 1} in an interleaved fashion~\cite{radhakrishnan2014senic, gulati2001nic, pratt2001arsenic}.
The specific way packets are sent from the client does not influence per-message processing approaches, such as INEC-TriEC, where the full message has to be received anyway to start the encoding. However, in packet-processing settings, the interleaving of the packets allows intermediate storage nodes to work in parallel on different packets, enabling the overlapping of the encoding of following packets with the aggregation (at the parity node) of the already encoded ones.

\subsubsection{Intermediate encoding} We define two handlers that are executed according to the role played by the storage node in the data encoding: i.e., storing data or parity. In the first case, for each packet, the packet payload gets encoded with the selected RS scheme and $m$ new intermediate parity packets are sent to the respective parity nodes. For encoding, we use the Galois field $GF(2^8)$. While this requires the handlers to scan the payload byte per byte, it allows us to use $256\times256$-byte lookup table to implement fast Galois field multiplication. The table is copied into NIC memory at DFS-initialization time and is shared between all DFS handlers.

\subsubsection{Final parities}
\begin{figure}[t]
	\centering
    \vspace{-1.5em}
	\includegraphics[width=0.9\columnwidth, trim=0 0 5 0]{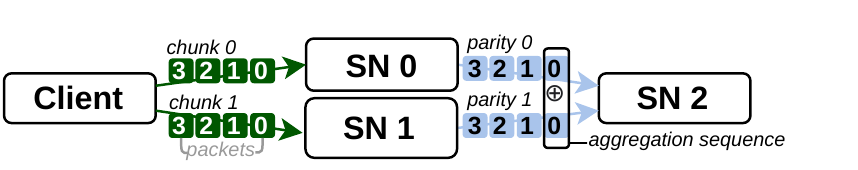}
	\vspace{-1em}
	\caption{RS(2,1) packets and aggregation sequence.}
	\vspace{-1.6em}
	\label{fig:ec_aggregation}
\end{figure}

Nodes selected to store parity chunks need to XOR the $k$ intermediate parity chunks to compute the final parity chunk: i.e., $\forall i \in [0, n): p_i^0 \oplus p_i^1 \oplus \dots \oplus p_i^{k-1}$, where $p_i^j$ indicates the $i$-th packet of the message carrying the intermediate parity computed by data node $j$. We define an \emph{aggregation sequence} as the sequence of packets $i$ coming from the $k$ data storage nodes: $\forall j \in [0, k): p_i^j$. 
Figure~\ref{fig:ec_aggregation} shows an example where the aggregation sequence for $i=0$ is marked:  
there, \emph{SN 2} must aggregate packets coming from both \emph{SN 0} and \emph{SN 1}, in the same order as they are produced. 

We keep a pool of accumulators (of size equal to the packet payload size) in the DFS state in NIC memory. 
The header handler tries to allocate an accumulator from the shared pool. If the pool is empty, then the aggregation cannot be done to the NIC and we fall back to a CPU-based aggregation. Otherwise, a mapping between the aggregation sequence ID $i$ and the accumulator is stored in an on-NIC hash table, allowing the subsequent payload handlers to target the same accumulator with atomic memory operations (i.e, XOR in this case).

Without the interleaved transmission of packets at the client, \emph{SN 0} and \emph{SN 1} would not be able to compute and send intermediate parities in parallel, delaying aggregation at \emph{SN 2} and negatively impacting the overall latency. Additionally, this would increase the time at \emph{SN 2} between the receiving of consecutive packets belonging to the same aggregation sequence, extending the allocation period of accumulator buffers.

\begin{figure}[h]
	\vspace{-1.5em}
	\centering
	\subfigure{\includegraphics[width=0.44\columnwidth, trim=0 0 5 0]{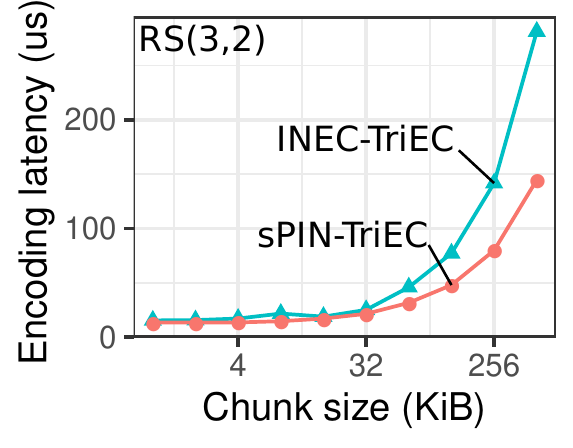}}
	\subfigure{\includegraphics[width=0.44\columnwidth, trim=0 0 5 0]{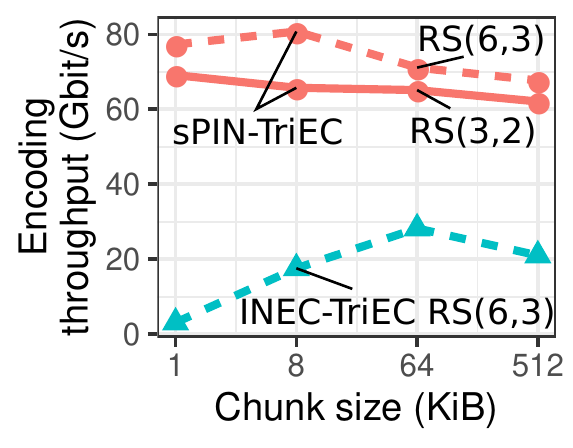}}
	\vspace{-1em}
	\caption{Encoding latency and throughput.}
	\vspace{-1.7em}
	\label{fig:ec_perf}
\end{figure}

\subsection{Erasure coding performance}

Unlike replication, where the sPIN handlers only forward packets to the next children in the broadcast tree, erasure coding needs to fully process packet data to encode it. This makes it challenging to achieve, e.g., 400 Gbit line rate where, with 2 KiB packets and 32 HPUs, each handler should not last more than $\sim$1310 ns to not become a bottleneck.

\paragraph{Encoding latency}
In Figure~\ref{fig:ec_perf} (left) we compare the write latency of sPIN-TriEC with the one of INEC-TriEC~\cite{shi2020inec}. INEC-TriEC operates on a per-chunk basis: at the intermediate storage nodes, chunks are first written into main memory, then read from the on-NIC EC accelerator to be encoded and sent to the parity nodes. With sPIN, we operate on a per-packet basis and encode packets on the fly without passing by the host's main memory. This allows sPIN-TriEC to have up to 2x lower latency. 
Since the TriEC results are taken from the INEC paper~\cite{shi2020inec} where a 100 Gbit/s network is used for experiments, we scale our simulated network to the same bandwidth.

\paragraph{Encoding bandwidth}
The encode bandwidth is computed with the same methodology of the INEC paper and common to window-based messaging benchmarks, that is \textit{bandwidth = (size of generated data)/(elapsed time)}. 
Figure~\ref{fig:ec_perf} (right) shows the PsPIN-TriEC encoding bandwidth for RS(3,2) and RS(6,3). For comparison, we plot the encoding bandwidth of INEC-TriEC for RS(6,3). 
sPIN-TriEC achieves up to 29x and 3.3x better bandwidth for 1 KiB and 512 KiB writes, respectively. For small block sizes, INEC-TriEC is penalized by memory copy overheads, which get better amortized with larger blocks. The sPIN-TriEC bandwidth does not directly depend on the block size because it always operates on packets but still experiences a 12\% drop in the throughput from 1 KiB to 512 KiB blocks. This is caused by the higher system utilization (i.e., more packets) that leads to more contention on NIC memory.

\begin{figure}[t]
	\vspace{-1.5em}
	\subfigure{{\includegraphics[width=0.59\columnwidth]{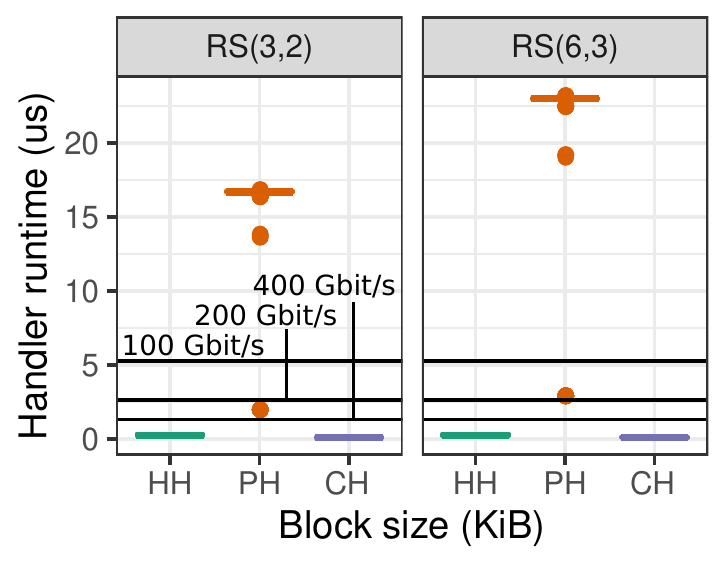} }}%
	\subfigure{{\includegraphics[width=0.41\columnwidth]{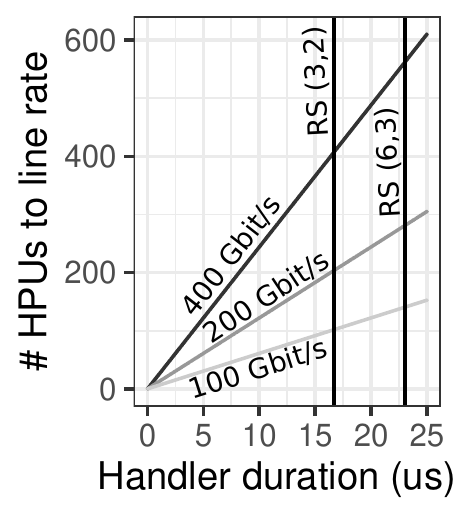} }}%
	\vspace{-1em}
	\caption{Left: handler running times for RS(3,2) and RS(6,3). Right: Number of HPUs needed to sustain 400 Gbit/s and 200 Gbit/s (2 KiB packets) for different average handler duration (x-axis).}%
	\vspace{-1.7em}
	\label{fig:ec_handlers}%
\end{figure}

\begin{table}[b]
	\scriptsize
	\vspace{-2em}
	\setlength{\tabcolsep}{1.5pt}
	\begin{tabularx}{\columnwidth}{p{1.5cm}XXXXXXXXX}
		\multirow{2}{*}{\textbf{Type}}  & \multicolumn{3}{c}{\textbf{Duration (ns)}} & \multicolumn{3}{c}{\textbf{Instructions}} & \multicolumn{3}{c}{\textbf{IPC}} \\
		& HH & PH & CH                               & HH & PH & CH                              & HH & PH & CH                     \\
		\midrule
		\textbf{RS(3,2)} 		& 215 & 16681 &  105 &    120 & 11672 & 35    & 0.56 & 0.7  & 0.33 \\
		\textbf{RS(6,3)}      	& 215 & 23018 &  82  &    120 & 16028 & 35    & 0.56 & 0.7  & 0.43 \\
		\bottomrule
	\end{tabularx}
	\vspace{0.5em}
	\caption{Handler statistics for different EC strategies.}
	\vspace{-2em}
	\label{table:ec_handlers}
\end{table}

\paragraph{Handlers analysis} Figure~\ref{fig:ec_handlers} (left) shows handler running times for RS(3,2) and RS(6,3). 
The horizontal lines show the per-handler time budget to sustain different network speeds.
Outliers are due to smaller payload packets carrying additional payload (i.e., when the MTU minus the packet header does not divide the block size). The running time of the payload handlers is dominated by the encoding loop, which goes over all the bytes of the packet payload and issues 5 instructions per byte for RS(3,2) and 7 for RS(6,3). 
Table~\ref{table:ec_handlers} reports statistics about handlers running times, number of instructions, and instructions per cycle (IPC). As these handlers are data-intensive, they do not sustain line rate on the selected PsPIN configuration. Figure~\ref{fig:ec_handlers} (right) shows how many HPUs are needed to sustain different network speeds given an average handlers duration time (x-axis). For example, the figure shows how for RS(6,3), a PsPIN configuration with 512 HPUs would allow sustaining 400 Gbit/s line rate for these handlers. Assuming to have a storage backend able to ingest this bandwidth, the modular architecture of PsPIN can be scaled out to sustain these types of workloads at line rate. 
For example, by increasing the number of clusters in PsPIN, we can increase the number of HPUs without introducing additional load on the per-cluster memories (L1).

\section{Discussion}

\noindent\emph{What if sPIN handlers do not run at line rate?} 
To sustain line rate, handlers must process packets within a limited time budget, depending on the line rate, the packet size, and the number of cores. If this is not the case, we have the same scenario as a receiver not being able to receive (i.e., to process in our case) fast enough. This can be mitigated by applying back pressure in the network~\cite{ieee802.1, reinemo2006overview} or by dropping packets.

\vspace{0.5em}\noindent\emph{How is data consistency (e.g., concurrent writes) handled?} 
Clients and metadata services coordinate to guarantee data consistency and avoid that, e.g., multiple clients write to the same part of a file. This coordination phase is part
of the control plane, while the actual data access is a part of the data plane. This separation of concerns is typical of several DFSs. Ceph~\cite{weil2006ceph} adopts the concept of capability that gives access rights (e.g., write) to clients and is granted by the management servers. To be able to issue a write, a client must first obtain the respective capability, eventually triggering its revocation from a client currently holding it. Similarly, HDFS~\cite{borthakur2008hdfs} clients need to be granted permission to write by the name nodes. As this work focuses on the offloading of DFS policies in the data plane, we do not assume any specific coordination protocol.

\vspace{0.5em}\noindent\emph{What happens if packets are lost?}  
We assume a lossless network where packet re-transmission (e.g., in case of data corruption) is performed directly by the NIC. In particular, we assume that once a packet is injected into the on-NIC PsPIN accelerator, the NIC already verified that the packet is not corrupted and that is not a re-transmission. This assumption is satisfied by modern lossless RDMA interconnects~\cite{sensi2020indepth, ib200}.

\vspace{0.5em}\noindent\emph{What happens if a storage node fails?}
A storage node that fails will not send acknowledgments to clients. A client that does not receive an acknowledgment for an ongoing operation after a predefined time threshold can start communicating with the metadata service to signal the failure and start the recovery process. The specific way the recovery is handled is DFS-dependent and not within the scope of this paper. 

\vspace{0.5em}\noindent\emph{What happens if a client fails?}
A client that fails while performing a write operation can leave some dangling state in the NIC of the storage nodes (e.g., \texttt{req\_table} in Listing~\ref{lst:handlers}). We extend PsPIN to associate a cleanup handler with each offloaded execution context. The cleanup handler is triggered by the PsPIN scheduler after an incoming message (i.e., a message for which the header packet has been received but the completion packet is still to come) is inactive for a specified amount of time. For the DFS execution context, the cleanup handler cleans any dangling state and generates an event on the storage node, signaling that a client write has been interrupted and allowing the DFS software to handle the client failure. 

\vspace{0.5em}\noindent\mbox{\emph{How to offload complex protocols?}}
DFSs can implement complex protocols and need large data structures to operate. For example, consensus protocols\mbox{\cite{lamport2019part, ongaro2015raft}} perform activities such as leader election, log replication, and sharding. While we do not advocate for the offloading of the full DFS logic to the NIC, we note that consensus protocols have been accelerated by extending RDMA primitives~\mbox{\cite{rdma-rsm}}. As sPIN provides an RDMA+X paradigm, these primitives can easily be implemented as sPIN handlers, delivering performance benefits without waiting for a their vendor-specific implementation.

\vspace{0.5em}\noindent\mbox{\emph{Offloading DFS building blocks in the cloud.}}
	One challenge of deploying sPIN in the cloud is handling fairness and quality-of-service (QoS) of NIC computing resources for multiple tenants. 
	While there is no multitenancy to enforce in disagraggated storage systems employing dedicated storage nodes, it is necessary to guarantee fairness and QoS in systems exploiting client-local persistent memories. 
	In these systems, network offloading is even more important due to (1) higher operating-system noise and (2) the need to reserve CPU time to dedicate to other tenants.

\begin{table}[h]
	\scriptsize
	\vspace{-0em}
	\setlength{\tabcolsep}{1.5pt}
	\begin{tabular}{l|c|ccc|p{2.5cm}}
		\textbf{DFS}       & \textbf{RDMA} &  & \textbf{Policies} &    & \textbf{Notes} \\
		&   & \textbf{Aut.} & \textbf{Rep.} & \textbf{EC}   &  \\
		\midrule
		\textbf{Lustre}~\cite{braam2019lustre}              & \faThumbsOUp  & \faThumbsUp & \faTimes      & \faTimes      & RPC+RDMA \\
		\textbf{IBM Spectrum Scale}~\cite{schmuck2002gpfs}  & \faTimes      &\faThumbsUp & \faThumbsUp   & \faThumbsUp   & \\
		\textbf{BeeGFS}~\cite{heichler2014introduction}     & \faThumbsOUp  &\faThumbsUp & \faThumbsUp   & \faTimes      & RDMA compatible \\
		\textbf{Ceph}~\cite{weil2006ceph}                   & \faTimes      &\faThumbsUp & \faThumbsUp   & \faThumbsUp   & \\
		\textbf{HDFS}~\cite{borthakur2008hdfs}              & \faThumbsOUp  &\faThumbsUp & \faThumbsUp   & \faThumbsUp   & RPC+RDMA~\cite{islam2012high}\\
		\textbf{Intel DAOS}~\cite{liang2020daos}            & \faThumbsOUp  &\faThumbsUp & \faThumbsUp   & \faThumbsUp   & RPC+RDMA \\
		\textbf{MadFS}~\cite{lu2009madfs}                   & \faThumbsUp   &\faThumbsUp & \faTimes      & \faTimes      &  \\
		\textbf{WekaIO Matrix}~\cite{wekaio}                & \faThumbsUp   &\faThumbsUp & \faTimes      & \faThumbsUp   &  \\ %
		\textbf{PanFS}~\cite{wang2014parallel}              & \faThumbsOUp  &\faThumbsUp & \faTimes      & \faThumbsUp   & RPC+RDMA \\
		\textbf{OrangeFS}~\cite{ross2000pvfs}               & \faThumbsOUp  &\faThumbsUp & \faThumbsUp   & \faTimes      & RPC+RDMA~\cite{wu2003pvfs} \\
		\textbf{Gluster}~\cite{davies2013scale}				&  \faThumbsOUp &\faThumbsUp &  \faThumbsUp   &  \faThumbsUp  & \\
		\textbf{Orion}~\cite{yang2019orion}                 & \faThumbsUp   &\faTimes & \faThumbsUp   & \faTimes      & Client-based replication.\\
		\textbf{Octopus}~\cite{lu2017octopus}               & \faThumbsOUp   &\faThumbsUp & \faTimes      & \faTimes      & RPC+RDMA\\
		\textbf{FileMR}~\cite{yang2020filemr}				& \faThumbsUp       &\faThumbsUp &  \faThumbsUp  & \faTimes     &  \\
		\bottomrule
	\end{tabular}
	\vspace{1em}
	\caption{DFS characteristics. \textbf{RDMA}: support for RDMA. \textbf{Aut.}: client authentication. \textbf{Rep.}: replication. \textbf{EC}: erasure coding. \faThumbsUp: provided, \faThumbsOUp: partially provided, \faTimes: not provided.}
	\vspace{-3.5em}
	\label{table:survey}
\end{table}

\section{Related work}
Table~\ref{table:survey} surveys state-of-the-art DFS systems, focusing on two main characteristics: RDMA support and use of different policies (client authentication, resilience via data replication, and resilience via erasure coding).

\noindent\textbf{RDMA- and SmartNIC-enabled storages.} 
With the evolution of networking and storage technologies, classical software storage stack with the 
operating system and host CPU in the loop have become bottlenecks. This led many distributed file systems to employ RDMA~\cite{braam2019lustre, schmuck2002gpfs, borthakur2008hdfs, liang2020daos, wang2014parallel, ross2000pvfs, lu2017octopus, yang2019orion, yang2020filemr}, and let data flow from storage nodes to clients and vice versa without CPU or OS intervention. 
One-sided RDMA operations are preceded by an RPC communication that validates the file access request and exposes the interested memory region over RDMA. Other approaches~\cite{yang2019orion, yang2020filemr}, utilize RDMA one-sided operations directly, relying on RDMA protection mechanisms. 

LineFS~\mbox{\cite{kim2021linefs}} accelerates DFS by exploiting NVIDIA BlueField NICs~\mbox{\cite{bluefield}}. Differently from LineFS, we target an event-based architecture (i.e., the event is the packet arrival) that is generally simpler (no hardware caches, simple RISC-V cores) but more parallel and specialized for packet processing. Being a DPDK-based approach, LineFS does not implement the user-level principle described in Section~\mbox{\ref{sec:principles}}.

1RMA~\mbox{\cite{singhvi20201rma}} proposes a data-center-optimized version of RDMA. They point out that in datacenter-scale storage systems the traditional connection-oriented RDMA approach can face scalability challenges. 
Our approach is orthogonal to 1RMA as it does not rely on long-lived RDMA connections.

In iPipe~\mbox{\cite{liu2019ipipe}}, an actor-based framework that schedules tasks between SmartNIC and host CPU, the authors discuss a replicated key-value store implementation based on the RDMA-Flat approach discussed in Section~\mbox{\ref{sec:replication}}.

\noindent\textbf{Network-offloaded data replication.}
Hyperloop~\cite{kim2018hyperloop} implements a ring-replication algorithm that can be offloaded to RDMA-capable NICs. It exploits triggered communication offered by Mellanox NICs, that can be used to express happen-before dependencies between pre-posted RDMA operations. Since pre-posted RDMA work requests do not depend on the content of the incoming message that triggered it, a client needs to configure them by remotely writing to their descriptor with an RDMA write, thereby configuring the broadcast ring. Once requests are configured, a client can start the offloaded replication by triggering requests on the first node of the ring. 
Tailwind~\cite{taleb2018tailwind} implements RDMA-accelerated replication that targets monotonically growing logs and delegates the replication process to the primary storage node, which can then use RPC+RDMA to communicate with replica nodes.

\noindent\textbf{Network-offloaded erasure coding.}
TriEC~\cite{shi2019triec} proposes a new EC NIC offload strategies that overcome many limitations of current-generation NIC-offloaded schemes for EC. INEC~\cite{shi2020inec} introduces a set of network primitives to accelerate NIC offloaded EC schemes that, similarly to Hyperloop, rely on pre-posted triggered communications that allow reducing EC encoding and decoding latencies.

\section{Conclusion}
We show how fully programmable SmartNICs fill the gap between full-RDMA solutions, which provide the best performance for one-sided accesses but do not expose enough compute capabilities to implement DFS policies, and CPU-based solutions, where DFS policies can be fully expressed at the cost of additional memory or network latencies.
Moreover, we show how on-NIC packet processing techniques can accelerate replication and erasure coding policies without requiring CPU or OS intervention. 
These results also demonstrate how, by having fully-programmable SmartNICs, fundamental DFS components can be offloaded to the network without depending on vendor-specific features and deployments.

All in all, these approaches to offload DFS policies can be followed by next-generation DFSs interfacing with fast storage media, where minimizing operation latencies (including policy enforcement) will be fundamental to minimize I/O overheads.

\section*{Acknowledgments}
This work has been partially funded by the European Projects \mbox{RED-SEA} (grant no.~955776) and \mbox{DEEP-SEA} (grant no.~955606). Daniele De Sensi is supported by an ETH Postdoctoral Fellowship (19-2 FEL-50).

\bibliographystyle{IEEEtran}
\bibliography{dfs}

\end{document}